\documentclass{emulateapj}
\usepackage{times}

\shorttitle{On the multiplicity of the ZAMS O star Her\,36}
\shortauthors{Arias et al.}

\begin{document}


\title{On the multiplicity of the zero-age main-sequence O star Herschel~36}


\author{Julia I. Arias$^1$,  Rodolfo H. Barb\'a$^{1,2}$, Roberto
  C. Gamen$^{3}$, Nidia I. Morrell$^{4}$, Jes\'us Ma\'{\i}z 
Apell\'aniz$^{5, 8}$, 
Emilio J. Alfaro$^{5}$,  Alfredo Sota$^{5}$, Nolan  R. Walborn$^{6}$ and
Christian Moni Bidin$^{6}$ }

\address{$^{1}$ Departamento de F\'{\i}sica, Universidad de La Serena,
Benavente 980, La Serena, Chile}

\address{$^{2}$ Instituto de Ciencias Astron\'omicas, de la Tierra y del Espacio
(ICATE-CONICET), Av. Espa\~na 1512 Sur, J5402DSP, San Juan, Argentina}

\address{$^{3}$Instituto de Astrof\'{\i}sica de La Plata (IALP-CONICET; 
Facultad de Ciencias Astron\'omicas y Geof\'{\i}sicas, Universidad Nacional 
de La Plata, Paseo del 
Bosque S/N, La Plata, Argentina}

\address{$^{4}$Las Campanas Observatory, The Carnegie Observatories, 
Casilla 601, La  Serena, Chile}

\address{$^{5}$Instituto de Astrof\'{\i}sica de Andaluc\'{\i}a-CSIC, Glorieta 
de la Astronom\'{\i}a s/n, Granada 18008, Spain}

\address{$^{6}$Space Telescope Science Institute\altaffilmark{9}, 
3700 San Martin Drive, Baltimore, MD 21218, USA}

\address{$^{7}$Departamento de Astronom\'{\i}a, Universidad de Concepci\'on,
Casilla 160-C, Concepci\'on, Chile}


\altaffiltext{8}{Ram\'on y Cajal Fellow}
\altaffiltext{9}{STScI is operated by AURA, Inc., under NASA contract 
NAS 5-26555}


\begin{abstract}
\noindent We present the analysis of high-resolution optical spectroscopic observations 
of the zero-age main-sequence O star Herschel~36 spanning six years. 
This star is definitely a multiple system, with at least three components 
detected in its spectrum. 
Based on our radial-velocity (RV) study, 
we propose a picture of a close massive  
binary  and a more distant companion, most probably in wide orbit about 
each other.
The orbital solution for the binary, whose components we identify as O9\,V 
and B0.5\,V, is characterized by a period of $1.5415\pm0.0006$ days.
With a spectral type O7.5\,V, the third body is the most luminous component of
the system and also presents RV variations with a period close 
to 498 days. 
Some possible hypotheses to explain the variability are briefly addressed and 
further observations are suggested.
\end{abstract}


\keywords{binaries: close --- stars: early-type ---  
stars: individual (Herschel~36)}



\section{Introduction}
\label{intro}

   The O-type star Herschel~36 (Her\,36) is located in the M8 Nebula, 
   a high-mass star-forming region at a distance of 1.3~kpc 
   (Arias et al. 2006). This O7.5\,V(n) star (Walborn 1982)
   is responsible for the ionization of the optically brightest part of M8, 
   known as the Hourglass Nebula.   
   A very young infrared cluster is found around Her\,36 (Arias et al. 2006). 
   The ultracompact H\,{\sc ii} region G5.97-1.17, a candidate ``proplyd'' 
   similar to those seen in the Orion Nebula (Stecklum et al. 1998), is 
   located $2\farcs7$ distant from Her\,36. Recently, high-resolution infrared 
   imaging of the region also revealed the existence of a compact source at 
   $0\farcs25$ southeast of the star, which harbors an early B-type embedded 
   star (Goto et al. 2006). Her\,36 and its
   close neighbors are probably part of a Trapezium-like stellar system.

   It seems clear that most, or even all, massive stars form in clusters and
   OB associations (de Wit et al. 2005). After a few hundred thousand years, 
   they become optically visible. 
   Only few, if any, of the massive stars are found as single objects and, on
   average, massive stars have more than one companion. Ignoring the
   multiplicity of massive stars can introduce serious problems when comparing
   observations and stellar evolution theory. 

   Although it is often stated that zero-age main-sequence (ZAMS) O stars
   should not be and are not observed optically, Her\,36 appears as a possible 
   bona fide ZAMS O star (Walborn 2007), based on the appearance of its 
   optical spectrum, the association with a dense, dusty nebular knot and its
   observed subluminosity (Arias et al. 2006). 
   Because of all these aspects, Her\,36 and the Hourglass have been many
   times compared to $\theta^{1}$~Ori~C and the Orion Trapezium.       
   Her\,36 has also been extensively studied because it shows an ``anomalous''
   extinction curve with $R_V=5.36$ (Arias et al. 2006), which represents one 
   of the highest $R_V$ values known (cf. Fitzpatrick \& Massa 2009). 

   A long-term spectroscopic monitoring of Galactic O and Wolf-Rayet of the
   Nitrogen sequence (WN) stars is being 
   carried out (Gamen et al. 2008), with the main 
   aim of establishing the multiplicity status of these objects. 
   As part of this observational campaign, Her\,36 has been systematically 
   observed. Here we present the results of the analysis of the data, which 
   demonstrate conclusively that Her\,36 is a complex multiple system, with 
   outstanding characteristics.

\section{Observations and data reduction}


Table~\ref{tbl-1} presents the journal of the observations used in this work. 
Most of them came from the monitoring 
campaign of Galactic O and WN stars and were obtained between 2005 April and 
2009 May at the observatories of Las Campanas (LCO) and La Silla, in Chile.
A total of 33 high-resolution spectra were obtained at LCO;
32 of those were taken with the \'echelle spectrograph at the 
2.5 m Du Pont telescope and the other with the MIKE at the 6.5 m 
Magellan II (Clay) telescope.
Twenty-six more spectra were acquired with FEROS at ESO La Silla. 
Four additional spectra were taken at the Complejo Astron\'omico El Leoncito
(CASLEO), Argentina, in observing runs prior to 2005, using the REOSC SEL
Cassegrain spectrograph in cross-dispersion mode.
Technical details of the observations and data reduction are presented by
R. H. Barb\'a et al. (2010, in preparation).


\section{The optical spectrum}

Besides the Balmer series, the spectrum of Her\,36 shows many absorption 
lines characteristic of hot massive stars, such as those of He\,{\sc i}, 
He\,{\sc ii}, C\,{\sc iii} and {\sc iv}, O\,{\sc iii} and Si\,{\sc iv}. 
Some diffuse interstellar bands at 4430, 4500, 5780, and 5797\,\AA, 
as well as a number of interstellar absorption lines due to Ca\,{\sc ii} and 
Na\,{\sc i}, are also evident. 
Narrow strong nebular emission lines, associated with the Hourglass Nebula, 
are observed, some of them superimposed on the Balmer and 
He\,{\sc i} absorptions. 

The spectrum of Her\,36 was monitored during two to six consecutive nights 
in each of the observing runs. The first result that came from the inspection 
of our high-resolution data is that the spectrum of this star is highly 
variable. 

The spectral lines of Her\,36 show large variations in both line-profile 
shape and radial velocity (RV), on a timescale of hours. 
Even if all the absorption lines appear appreciably broadened, 
in good agreement with the qualifier (n) in the spectral classification 
by Walborn (1982), the profiles of the most prominent ones alternate between  
more or less normal and unusually broad. 
When observed with enough resolution, the lines are clearly not single: 
at different times and in different lines, one, two, or three components are
detected, showing that this star is actually a multiple system.

\begin{figure*}
\epsscale{.90}
\plotone{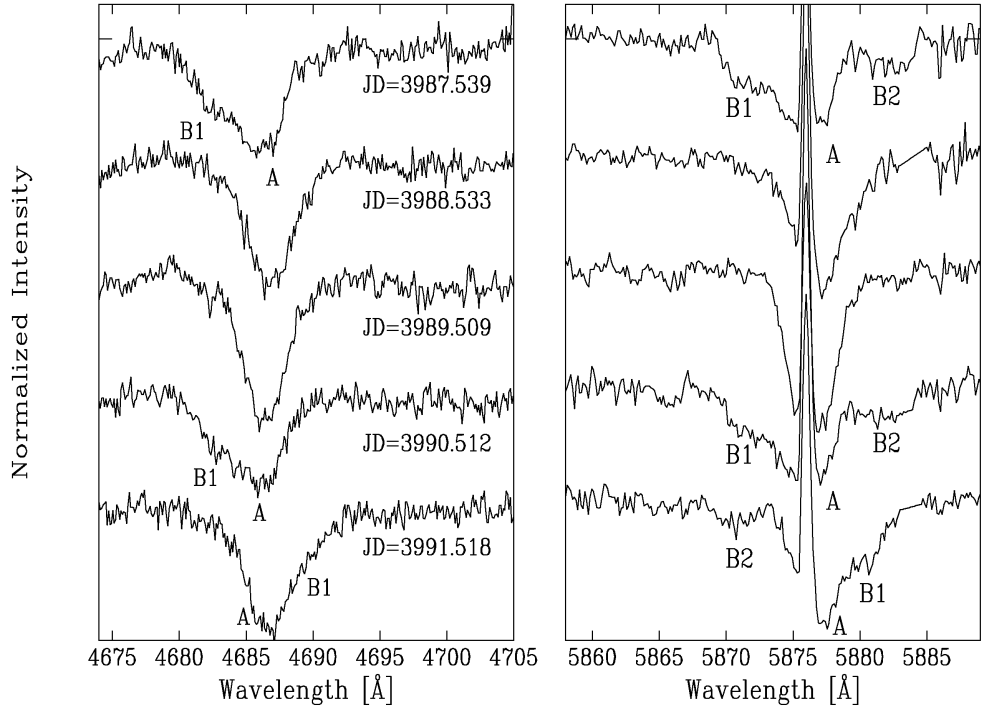}
\caption{Variations in the He\,{\sc ii}~$\lambda$4686 (left panel) and 
He\,{\sc i}~$\lambda$5875 (right panel) line profiles, observed in LCO/Du Pont 
spectrograms obtained in five consecutive nights. 
The spectra are labeled with the corresponding Julian Date (JD) in the format 
HJD-2,450,000.
All the spectra are normalized to unity and have equal scales. 
The separation between continuum levels is 10\% of the normalized flux.
A, B1, and B2 refer to the components described in the text.
The strong emission line superimposed on the He\,{\sc i} absorption is 
associated with the Hourglass Nebula.}
\label{reg4600}
\end{figure*}

We present the variations in the He\,{\sc ii}~$\lambda$4686 line
in the left panel of Figure~\ref{reg4600}. This plot includes the data from 
five consecutive nights. The morphology observed 
varies from a very deep single profile to a well-defined double line. 
At first sight, however, the line-profile morphology does not seem to be 
strictly repeatable. Large RV variations accompany all these changes. 
We will refer to the main and secondary absorptions as ``component A'' and 
``component B1'', respectively. 

An even more complex behavior is observed in the He\,{\sc i} absorption lines. 
The right panel of Figure~\ref{reg4600} shows the spectral variations in the 
He\,{\sc i} $\lambda$5875 line. Now three absorption components (labeled as 
A, B1, and B2) can be clearly identified. 
Whereas components A and B1 correlate 
with those observed in the He\,{\sc ii}~$\lambda$4686 profile, the component 
B2 is undetected in the He\,{\sc ii} lines, indicating that the latter star 
is cooler than its companions.
The RV of the main component A changes rather slowly, but the 
variations observed for components B1 and B2 are remarkably large.
Note that the positions of the components B1 and B2 practically invert 
from the forth to the fifth night. 

Apart from the He\,{\sc ii}~$\lambda$4686 and He\,{\sc i}~$\lambda$5875
lines, several other photospheric features, such as He\,{\sc i}~$\lambda$4471, 
He\,{\sc ii}~$\lambda$5411, C\,{\sc iii}~$\lambda$4650, and
C\,{\sc iv}~$\lambda$5812, were also measured.
We found correlated variations in profile shape and RV in most of these lines. 

Finally, we considered the least blended spectra in order to obtain a spectral 
type for each of the three stellar components detected in the
spectrum of Her\,36. The spectral classification was performed by comparison 
of the spectrum with the OB star atlas of Walborn \& Fitzpatrick (1990).
We adopt the spectral types O7.5\,V, O9\,V, and B0.5\,V for components A, B1,
and B2, respectively.

\section {A picture for the multiple system Her~36}
\label{period}

The fact that all measured photospheric lines show similar profile and RV 
variations suggests that the RV variations detected in the spectrum of 
Her\,36 are actually caused by the orbital motion of, at least, three stellar 
components. We propose that the system is composed of a close O-type binary 
and a third more distant companion, also of spectral type O. 
The binary and the third body are very likely in wide orbit about each other.

\subsection{The Close Pair B1+B2}
\label{pairB}

The behavior observed in the He\,{\sc i}~$\lambda$5875 line suggests that 
the components B1 and B2 are constituents of a double-lined spectroscopic 
binary. When we refer to this binary system we will simply use the term 
``system B''.

As mentioned before, the component B2 is not detected in He\,{\sc ii}. 
Moreover, only in the He\,{\sc i}~$\lambda$5875 line, the profile features 
appear separated enough to obtain a reliable measurement of the RV of each 
component. Thus we considered only this line to determine the orbital solution 
for the stellar system B. 
The RV measurements used in the determination of the orbital solution are 
presented in Table~\ref{tbl-1} (Columns 6 and 7). 
We determined the RVs of the different components of the He\,{\sc i} 
$\lambda$5875 line by fitting them with adequate sums of Gaussian profiles. 
To do this, we considered four Gaussian functions: one for each of the stellar 
components A, B1, and B2, as well as an additional one to take into account 
the strong nebular emission.
First, we used the least blended spectra to determine the intensities 
and widths of the three stellar blends and then repeated the disentangling
by keeping these parameters fixed. Thus the deconvolution of Gaussian
profiles yielded the central wavelengths of the four components (three stellar 
and one nebular) of the He\,{\sc i}~$\lambda$5875 line as adjusted parameters.
In spite of the complexity of this blend, the fits were highly satisfactory 
in most of the cases. Being conservative, we estimate that the measurement 
errors in the RVs of the stellar components are no larger than 10~km\,s$^{-1}$.

\begin{table*}
\centering
\caption{Radial-velocity measurements of the stellar components observed 
in the spectra of Her\,36}
\label{tbl-1}
\begin{tabular}{ccccccccc}
\hline\\
No. &HJD & \multicolumn{2}{c}{He\,{\sc ii}~4686}  &  \multicolumn{3}{c}{He\,{\sc i}~5876}   & Telescope & Instrument \\
&-2450000        & A           & B1   &  A         & B1  & B2 &           &  \\
\hline\\
1 & 2471.6791 & -10.2 &  126.0 & ...   & ...    & ...     & Casleo-2.15 & REOSC  \\
2 & 2835.6685 & -58.8 &   92.9 & ...   & ...    & ...     & Casleo-2.15 & REOSC  \\
3 & 2836.7348 & -72.5 &  218.4 & -65.0 & 256.8  & -229.5  & Casleo-2.15 & REOSC  \\
4 & 2837.6846 & -54.5 & -146.9 & ...   & ...    & ...     & Casleo-2.15 & REOSC   \\
5 & 3490.8852 &  26.4 & -128.7 & ...   & ...    & ...     & LCO-DP      & Echelle   \\
6 & 3491.9145 &  13.6 &  205.3 & 16.6  & 213.8  & -336.1  & LCO-DP      & Echelle   \\
7 & 3491.9199 &	 19.3 &  210.5 & 18.6  & 231.4  & -338.8 & LCO-DP      & Echelle    \\
8 & 3772.8789 & -47.0 &   50.3 & ...   & ...    & ...     & LCO-DP      & Echelle   \\
9 & 3873.9131 & -50.9 &   91.2 & ...   & ...    & ...     & LCO-DP      & Echelle   \\
10& 3874.9028 & -44.9 & -175.5 & -46.9 & -132.5 & 347.3   & LCO-DP      & Echelle    \\
\hline\\
\end{tabular}
\tablecomments{Rest wavelengths used in the determination of the RVs
are 4685.65~\AA\, and 5875.65~\AA, respectively.\\
(This table is available in its entirety in a machine-readable form in the
online journal. A portion is shown here for guidance  regarding its form and
content.)}
\end{table*}

To derive the period of the orbital motion, we applied a variety of methods 
to the RV measurements corresponding to the components B1 and B2.
These routines include the Lafler \& Kinman (1965) method and 
subsequent modifications of it (Marraco \& Muzzio 1980, MM80), 
as well as information entropy minimization (Cincotta et al. 1995, 
CMN95). We considered three RV data sets: (1) the measurements of the 
primary component, B1, alone; 
(2) the measurements of the secondary component, B2, alone; and 
(3) the measurements of both components together.
Based on the resulting periodograms, we adopted $P=1.54$~days 
as the best estimate for the period.

\begin{figure}
\includegraphics[width=8cm, angle=0]{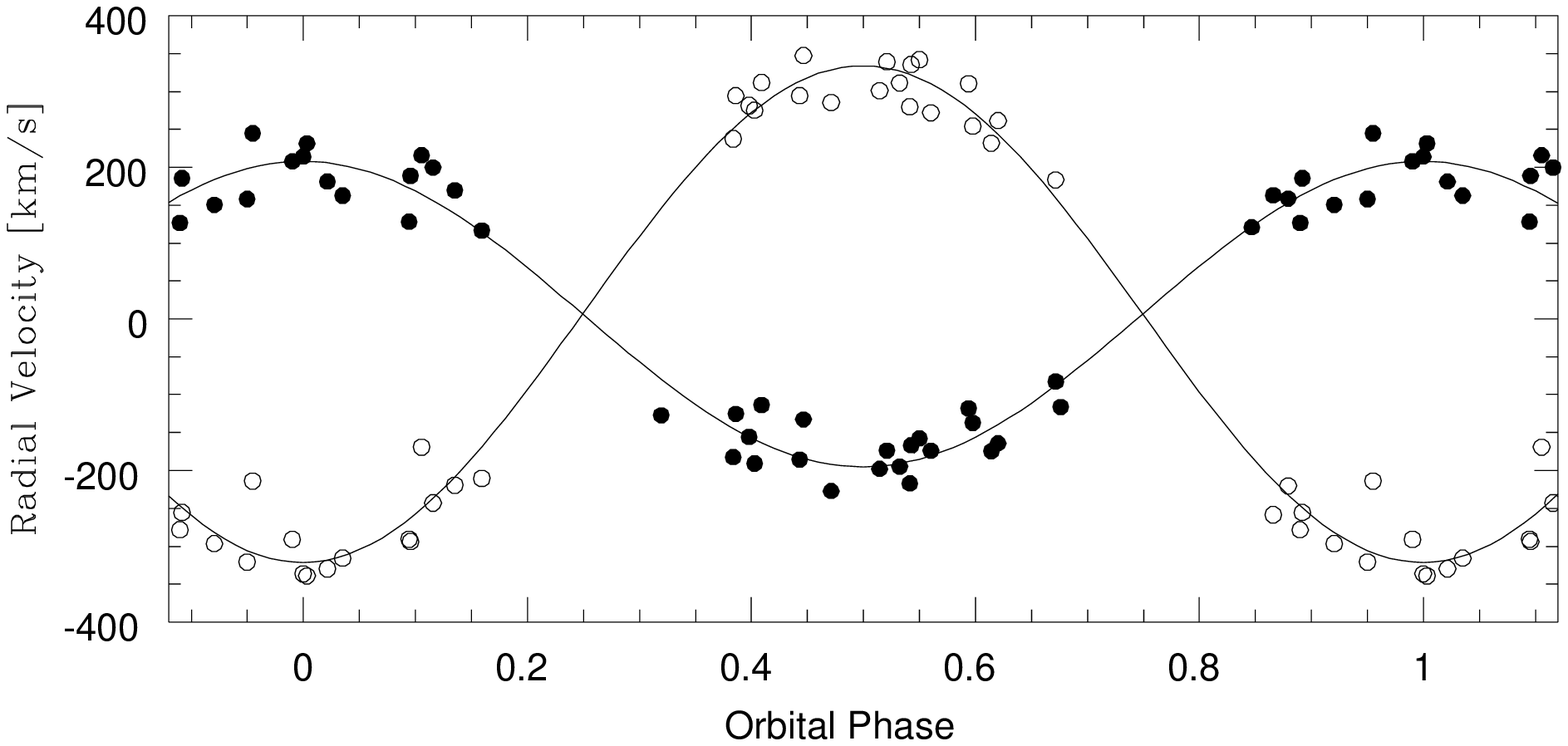}

\includegraphics[width=8cm, angle=0]{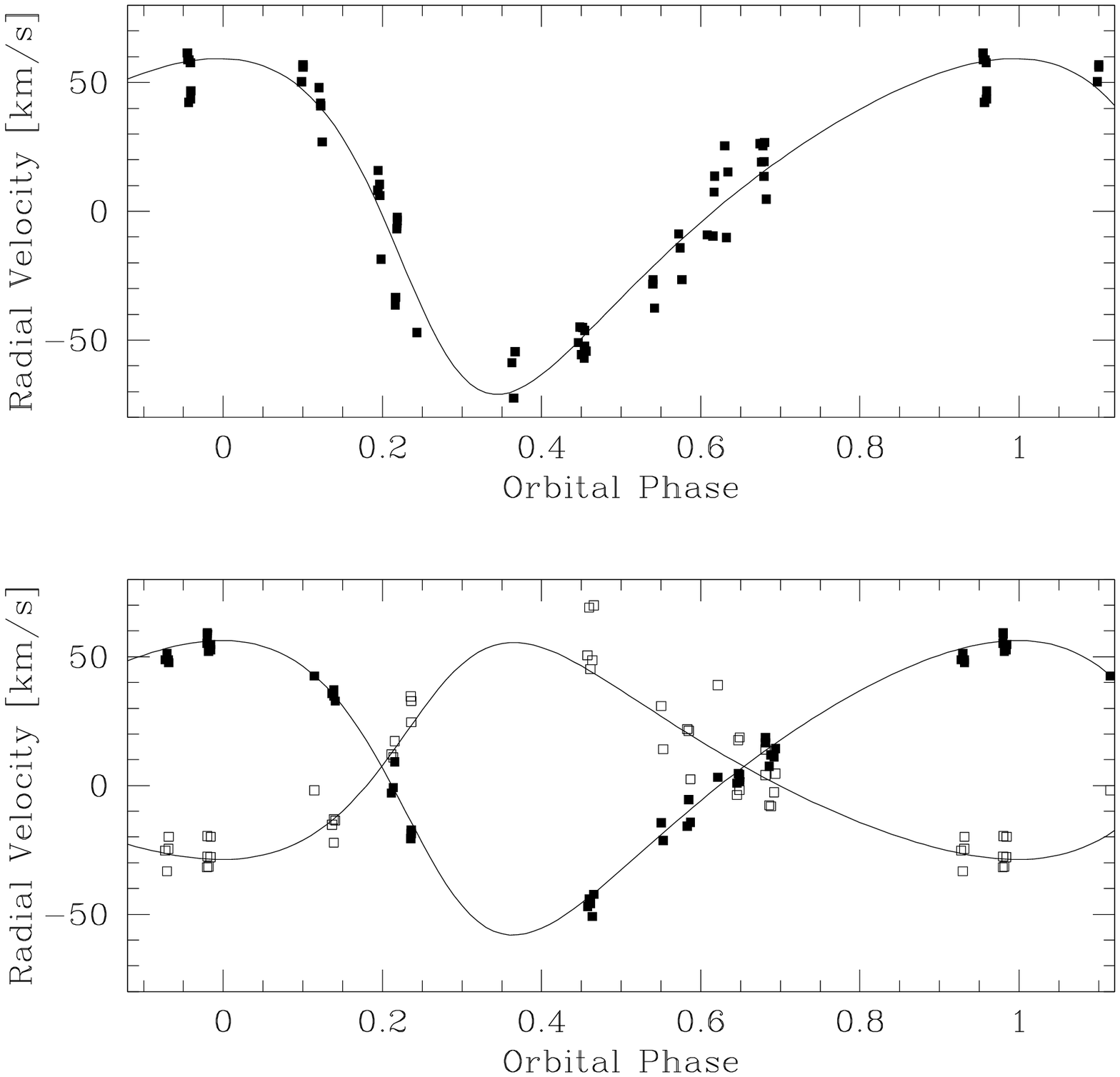}

\caption{Upper panel: RV curves corresponding to the close stellar pair B1+B2. 
The filled and open circles indicate the observed RVs of the primary (B1) and 
secondary (B2), respectively. Zero phase is a time of maximum velocity for the 
primary B1. Middle panel: the RV curve derived from the measurements of the 
strongest component (component A) of the He\,{\sc ii}~$\lambda$4686 line. 
Lower panel: RV curves corresponding to a joint orbital solution obtained from 
the He\,{\sc i}~$\lambda$5876 line, considering the RVs of the component A for 
the primary (filled squares) and the systemic velocities of the system B for 
the secondary (open squares). The orbital parameters that characterize these 
solutions are listed in Table~\ref{tbl-3}. 
Zero phase is a time of maximum velocity for the component A.
}
\label{RVcurves}
\end{figure}

To derive the orbital elements of the system B, we used a modified version of 
the code originally written by Bertiau \& Grobben (1969), adopting the value 
$P_0=1.54$ days as starting estimate of the period and considering 
the same data sets as previously described. After some trials with general
eccentric fits, a circular orbit was assumed for the system. 
The best orbital solution is presented in Table~\ref{tbl-2} and illustrated 
by the RV curves in the upper panel of Figure~\ref{RVcurves}.

\begin{table}
\begin{center}
\caption{Orbital and physical parameters of the system B\label{tbl-2}}
\begin{tabular}{lc}
\tableline
& \\
$P_S$ (days)                         & $1.5415\pm0.0006$ \\
$T_{\rm Vmax}$~(HJD-2,450,000)      & $4248.8053\pm0.0061$\\
$V_o$~(km\,s$^{-1}$)            & $6.3\pm3.6$\\
$K_1$~(km\,s$^{-1}$)           & $201.4\pm5.8$ \\
$K_2$~(km\,s$^{-1}$)           & $327.3\pm5.8$ \\
$a_1$\,{$\sin{i_B}$}~(km)         & $(42.7\pm1.2)\times10^5$ \\
$a_2$\,{$\sin{i_B}$}~(km)         & $(69.4\pm1.2)\times10^5$ \\
$M_1$\,{$\sin^3{i_B}$}~(M$_{\odot}$) & $14.6\pm1.3$ \\                  
$M_2$\,{$\sin^3{i_B}$}~(M$_{\odot}$) & $9.0\pm1.3$ \\
$Q~(M_2/M_1)$                  & $0.61\pm0.03$ \\
rms (km\,s$^{-1}$)             & 31.0 \\
& \\
\tableline
\end{tabular}
\tablecomments{Parameters derived from the RVs of the 
He\,{\sc i}~$\lambda$5875 line and assuming a circular orbit. $T_{\rm Vmax}$
refers to the maximum velocity of the primary star B1.}
\end{center}
\end{table}

\subsection{The Most Luminous Star A}

The star A is the most luminous component of the system and dominates the
optical spectrum of Her\,36. Although not so evident, it also presents 
RV variations. Careful inspection of the data suggests the presence of a 
long-term (few hundred days) variability. 

To study the spectral variations, we used the RV measurements of 
the He\,{\sc ii}~$\lambda$4686 line. This line  appears as one of the most 
useful features for investigating possible periodicities 
because of being less contaminated by other features or the background nebular 
emission, and, mainly, because of showing only two stellar components: A and
B1. As in the case of the He\,{\sc i}~$\lambda$5875 line, we determined the
RVs of the different components by performing multiple Gaussian fits. These
measurements are presented in Columns 3 and 4 of Table~\ref{tbl-1}. 
Due to the strong overlap of the line components, our initial disentangling 
attempts yielded systematic shifts of the central position of the
component A, which at first we erroneously interpreted as short-term periodic 
RV variations. However, this effect was fully corrected when we repeated the
disentangling by keeping the intensities and widths of the Gaussian profiles
fixed, in order that the wavelength positions of A and B1 were the only
parameters adjusted in the fit. As a result, we obtained very good quality 
fits and reliable RVs, with typical measurement errors no larger than
10~km\,s$^{-1}$.

To estimate the possible period of the variability of A, we applied the same 
period-finding routines as in Sec.~\ref{pairB}, i.e., the methods from MM80 
and CMN95, to the corresponding RV measurements, and we also performed 
the Lomb-Scargle (LS) periodogram search (Scargle 1982).
A number of possible values of the period can be identified from the 
periodograms peaks, the most relevant corresponding to $P=508\pm15$ days. 
An additional value of $P=206$ days is also strong but less significant, and 
cannot be discarded a priori. 

Considering the values from the periodogram peaks as starting estimates of 
the period, we explored the possibility of deriving an orbital solution
for the motion of the component A. For this purpose, we applied the same 
code as in Sec.~\ref{pairB} to the RVs of the component A of the 
He\,{\sc ii}~$\lambda$4686 line. The resulting solution is presented in the 
second column of Table~\ref{tbl-3} and illustrated by the RV curve in the 
middle panel of Figure~\ref{RVcurves}.
An independent solution based on the RV measurements of the component A of the
He\,{\sc i}~$\lambda$5876 line (Column 5 of Table~\ref{tbl-1}) 
was also calculated, and it is presented in the
third column of Table~\ref{tbl-3}. It can be verified that, within the errors, 
both solutions represent essentially the same orbit.\\

\section{Summary and discussion}

The main finding of this study is the multiple nature of the ZAMS object 
Her\,36. The absorption lines in the spectrum of this 
star are clearly not single. We arrived at the conclusion that Her\,36 
presents at least three stellar components that we labeled as A, B1, and B2, 
and classified as O7.5\,V, O9\,V, and B0.5\,V, respectively.

Arias et al. (2006) determined the distance and reddening to several
early-type stars in the Hourglass region, including Her\,36. 
As mentioned in Sec.~\ref{intro}, Her\,36 that was considered as a single 
O7.5\,V star in their calculus fell below the calibration of its luminosity 
class. The here demonstrated multiplicity of Her\,36 makes the stars that 
compose this triple system even much more subluminous individually. 
Thus, if we adopt the absolute magnitudes for ZAMS OB stars from 
Hanson et al. (1997), the combined $M_V$ value obtained is similar to
that quoted by Arias et al. (2006). This fact adds additional support to the 
hypothesis that the stellar components of this system may be on the ZAMS. 

We measured the most prominent features in the spectrum of Her\,36, finding 
correlated variations in profile shape and RV. Based on our RV study, we 
proposed that Her\,36 is composed of a close massive binary, the stellar pair 
B1+B2 (system B), and a companion, the component A, which is the most luminous 
star of the system and dominates the optical spectrum.

Considering the RV measurements of the He\,{\sc i}~$\lambda$5875 line, we 
derived a set of orbital elements for the system B. The best solution is 
characterized by a circular orbit with a period $P_S=1.5415\pm0.0006$ days, 
and semi-amplitudes $K_1=201.4$~km\,s$^{-1}$ and $K_2=327.3$~km\,s$^{-1}$. 
We can use the obtained minimum masses $M\sin^{3}i$ to estimate the 
inclination of the system B. The values of 14.6 and 9.0~M$_{\odot}$, derived 
for the primary and the secondary, respectively, are compared with the typical 
masses of O9\,V and B0.5\,V stars. According to the values for detached 
eclipsing systems from Gies (2003) (19.0~M$_{\odot}$ for O9\,V and 
11.7~M$_{\odot}$ for B0.3\,V), the system must have a large inclination (larger 
than 67$^{\circ}$), which means that it could possibly undergo eclipses. 
Unfortunately, no photometric study of Her~36 exists at present. Future 
photometric monitoring will certainly be valuable.

The component A also shows RV variations. Independent RV measurements of the
 He\,{\sc ii}~$\lambda$4686 and  the He\,{\sc i}~$\lambda$5876 lines yield 
essentially the same relatively eccentric ($e=0.3$) orbit, characterized by 
a period close to 498 days and a semi-amplitude of the order of 65~km\,s$^{-1}$.

We feel that the component A and the close binary system B are in wide orbit 
about each other. The long-term periodicity observed in the RVs of the 
component A must then be related to the mutual orbit of both objects. 
If that is the case, and we are indeed in the presence of a triple
(hierarchy 2+1) system, then we should expect the systemic velocity of the 
binary B to change with a period $P_L\approx498$ days. 
To prove this hypothesis, we computed a joint orbital solution considering 
the RVs of the component A of the He\,{\sc i}~$\lambda$5876 line for the 
primary, and the systemic velocities of the system B, $v_B$, for the 
secondary. The $v_B$ values were calculated according to the basic expression 
$v_s=(M_{1}v_{1} + M_{2}v_{2})/(M_1+M_2)$, where $M_{1}$ and $M_{2}$ are the
minimum masses from Table~\ref{tbl-2}, and $v_{1}$ and $v_{2}$ are the RVs of
the components B1 and B2 of the He\,{\sc i}~$\lambda$5876 line, respectively.
The resulting solution is characterized by the parameters in the last column 
of Table~\ref{tbl-3} and the RV curves in the lower panel of 
Figure~\ref{RVcurves}.
Although with somewhat larger dispersion, these curves represent practically 
the same orbit as those derived from the RV data of A alone, as can be easily
verified by confronting the orbital parameters listed in the different 
columns of Table~\ref{tbl-3}. We note here that the joint solution was 
obtained in a completely independent manner, supporting the idea of the mutual 
orbital motion between A and B.

Comparing the projected mass of the system B from Table~\ref{tbl-3} with the 
projected total mass of the close pair B1+B2 from Table~\ref{tbl-2}, it can be 
shown that $\sin{i_A}=1.03\sin{i_B}$. This means that the inclinations of the
two orbital planes satisfy either $i_B \simeq i_A$ or 
$i_B \simeq 180^{\circ}-i_A$. In the first case, the inner and the outer 
orbits of the triple system will be nearly coplanar, whereas in the last case 
the relative inclination between the orbital planes, $\phi$, will lie between 
$180^{\circ}-2\,i_A\leq\phi\leq180^{\circ}$ (Fekel 1981).
The relative orientation of the orbits in a multiple system is of particular 
interest because of its direct relationship with the conditions at the epoch 
of formation. Thus, the study of a triple system as young as Her\,36 can 
provide important clues to the formation processes of massive stellar systems.

On the other hand, comparing the minimum mass for the primary A 
(19.2~$M_{\odot}$) to typical masses of O7.5\,V stars from Gies (2003), 
we estimate that $i_A$ is most probably larger than 70$^{\circ}$. 
The component A has the largest individual mass of the three stars. Its mass 
is however fairly comparable with the total mass of the inner binary B.
According to Table~\ref{tbl-3}, the minimum mass of the triple system is 
$19.2+26.0=45.2$~$M_{\odot}$. Assuming a minimum inclination 
$i_A\simeq70^{\circ}$, we conclude that the total mass of Her\,36 is probably 
no larger than 54~$M_{\odot}$.

The triple ZAMS system Her\,36 is really peculiar, and it
undoubtedly deserves additional study. Clearly, further observations are 
required before a complete characterization of the orbits can be given 
and a definitive model for the system can be derived. Spectroscopic 
observations with higher signal-to-noise ratio, as well as an exhaustive program of 
photometric monitoring, seem essential.

\begin{table*}
\begin{center}
\caption{Orbital and physical parameters of the multiple system 
Her\,36\label{tbl-3}}
\begin{tabular}{lccc}
\tableline\tableline
& &\\
Parameter & He\,{\sc ii} $\lambda$4686 & He\,{\sc i} $\lambda$5876 &  He\,{\sc i} $\lambda$5876\\
& A & A & A+B \\
\tableline
& & &\\
$P_L$ (days)                   & $498.3\pm3.7$ & $496.3\pm2.2$ & $493.3\pm3.2$ \\
$e$                          & $0.29\pm0.04$ & $0.30\pm0.03$ & $0.25\pm0.03$ \\
$\omega$ (deg)           & $126.0\pm6.3$   & $126.9\pm3.4$ &  $124.5\pm5.4$\\          
$T_o$~(HJD-2,450,000)        & $4279.5\pm7.5$ & $4278.6\pm3.9$ &  $4275.6\pm6.5$\\
$T_{\rm Vmax}$~(HJD-2,450,000)   &  $4147.5\pm7.5$ & $4146.7\pm3.9$ &  $4141.5\pm6.5$\\
$V_o$~(km\,s$^{-1}$)         & $5.3\pm1.8$& $4.4\pm0.9$ &  $7.4\pm0.8$ \\
$K_A$~(km\,s$^{-1}$)          & $65.2\pm3.1$ & $61.2\pm2.2$ &  $57.2\pm2.4$ \\
$K_B$~(km\,s$^{-1}$)        &      -       &  -           & $42.1\pm2.4$ \\
$a_A$\,{$\sin{i_A}$}~(km)     & $(42.7\pm3.0)\times10^7$ & $(39.8\pm1.9)\times10^7$ &  $(37.5\pm2.1)\times10^7$\\
$a_B$\,{$\sin{i_A}$}~(km)     & - & - & $(27.6\pm2.0)\times10^7$ \\
$f(M)$~(M$_{\odot}$)         & $12.5\pm2.4$ & $10.2\pm1.5$ & - \\
$M_A$\,{$\sin^3{i_A}$}~(M$_{\odot}$) &  - & - & $19.2\pm6.1$ \\                  
$M_B$\,{$\sin^3{i_A}$}~(M$_{\odot}$) &  - & - &$26.0\pm6.3$ \\
$Q~(M_B/M_A)$                    & -  & -  & $1.36\pm0.13$ \\
rms (km\,s$^{-1}$)          & 10.3 & 3.6 & 7.0\\
& & & \\
\tableline
\end{tabular}
\tablecomments{Columns 2 and 3 present the orbital parameters derived from the 
RVs of the component A alone, considering the He\,{\sc ii}~$\lambda$4686 and 
the He\,{\sc i}~$\lambda$5876 lines, respectively. Column 4 presents a joint 
orbital solution obtained from the He\,{\sc i}~$\lambda$5876 line, considering 
the RVs of the component A for the primary and the systemic velocities of the 
system B for the secondary.}
\end{center}
\end{table*}


\acknowledgments

We would like to thank our anonymous referee whose critical remarks helped us
to substantially improve this Letter. 
J.I.A. and R.H.B. acknowledge financial support from DIULS through projects 
No. PI07101 and CD08102, respectively and JIA from Fondo ALMA para el 
Desarrollo de la Astronom\'{\i}a CONICYT No. 31050004. 
J.M.A., E.J.A. and A.S.  acknowledge support from the Spanish Government MICINN through
grant AYA2007-64052 and from Consejer\'{\i}a de Educaci\'on y Ciencia (Junta
de Andaluc\'{\i}a) through TIC-101 and TIC-4075. JMA acknowledges support 
from the Ram\'on y Cajal Fellowship program and FEDER funds.  
CMB acknowledges the Chilean Centro de Excelencia en Astrof\'isica y 
Tecnolog\'ias Afines (CATA). This work relies on data taken at the La Silla
Observatory under Program IDs 077.B-038(A), 079.D-0564(A), 079.D-0564(C),
081.D-2008(A), 081.D-2008(B), and 083.D-0589(A).

\clearpage




\begin{thebibliography}{}

\bibitem[Arias(2006)]{a06} Arias, J. I., Barb\'a, R. H., Ma\'{\i}z 
Apell\'aniz, J., Morrell, N. I., \& Rubio, M. 2006, \mnras, 366, 739

\bibitem[bertiau(1969)]{b69} Bertiau, F., Grobben, J. 1969,
  Ric. Astron. Sp. Vaticana, 8, 1

\bibitem[Cincotta(1995)]{c95} Cincotta, P. M., M\'endez, M., Nu\~nez, J. A. 1995, ApJ, 449, 231

\bibitem[de Wit(2005)]{dw05} de Wit, W. J., Testi, L., Palla, F., \&
  Zinnecker, H. 2005, \aap, 437, 247

\bibitem[Fekel(1981)]{f81} Fekel, F. C. 1981, ApJ, 246, 879

\bibitem[Fitzpatrick(2009)]{f09} Fitzpatrick, E. L., \& Massa, D. 2009, ApJ,
  699, 1209

\bibitem[Gamen(2008)]{g08} Gamen, R. C., Barb\'a, R. H., Morrell, N. I.,
  Arias, J. I., Ma\'{\i}z  Apell\'aniz, J.  2008, RevMexAAC, 33, 54

\bibitem[Gies(2003)]{g03} Gies, D. R. 2003, in IAU Symp 212, A Massive Star
  Odyssey: From Main Sequence to Supernova, ed. K. van der Hutch, A. Herrero,
  \& C. Esteban (Dordrecht: Kluwer), 91

\bibitem[Goto(2006)]{g06} Goto, M., Stecklum, B., Linz, H., Feldt, M., Henning
     T{\sc h}., Pascucci, I., \& Usuda, T. 2006, \apj, 649, 299

\bibitem[Hanson(1997)]{h97} Hanson, M. M., Howarth, I. D., \& Conti, P. S. 1997,
\apj, 489, 698

\bibitem[Lafler(1965)]{l65} Lafler, J., Kinman, T. D. 1965, ApJS, 11, 216

\bibitem[Marraco(1980)]{m80} Marraco, H. G., Muzzio, J. C. 1980, PASP, 92, 700

\bibitem[Scargle(1982)]{s82} Scargle, J. D. 1982, ApJ, 263, 835


\bibitem[Stecklum(1998)]{s98} Stecklum, B., Henning, T., Feldt, M., Hayward,
  T.L., et al 1998, AJ, 115, 767

\bibitem[Walborn(1982)]{w82} Walborn, N. R. 1982, AJ, 87, 1300

\bibitem[Walborn(1990)]{w90} Walborn, N. R., Fitzpatrick, E. L. 1990, 
PASP, 102, 379 

\bibitem[Walborn(2007)]{w07} Walborn, N. R. 2007, in Proc. STScI Symp, 20 
`Massive Stars: From Pop III and GRB to the Milky Way, ed. M. Livio \& E. 
Villaver (Cambridge: Cambridge Univ. Press), 167

\end{thebibliography}
\end{document}